\documentclass[usenatbib,referee]{mn2e}
\usepackage{graphicx}

\def\deg{\ifmmode^\circ\else$^\circ$\fi}

\def\msun{M$_{\odot}$}

\def\mic{\,$\mu$m}

\title[The sequence of low and high mass star formation in the young
stellar cluster I19343+2026]{The sequence of low and high mass star
formation in the young stellar cluster IRAS 19343+2026}

\author[Ojha, D.K., Kumar, M.S.N., Davis, C.J., and Grave, J.M.C.]
{D. K. Ojha$^{1}$\thanks{E-mail: ojha@tifr.res.in},
M. S. N. Kumar$^{2}$, C. J. Davis$^{3}$ and J. M. C. Grave$^{2}$
\\ $^{1}$ Tata Institute of Fundamental Research,
Homi Bhabha Road, Mumbai-400005, India\\ $^{2}$Centro de Astrof\'{\i}sica
da Universidade do Porto, Rua das Estrelas, 4150-762 s/n Porto,
Portugal \\$^{3}$Joint Astronomy Center, 660 N. A'oh\={o}k\={u} Place,
University Park, Hilo, HI 96720, USA}

\begin{document}

\date{ }

\pagerange{\pageref{firstpage}--\pageref{lastpage}} \pubyear{2010}

\maketitle

\label{firstpage}

\begin{abstract}

$BVRIJHK$ photometry, {\em Spitzer}-GLIMPSE photometry and $HK$
band spectroscopy were used to study the stellar content of IRAS
19343+2026, a (proto)star/cluster candidate, located close to the
Galactic plane. The data suggest that
IRAS 19343+2026 is a rich cluster associated with a massive protostar
of 7.6 \msun\ with
an age of $\sim$ 10$^{5}$ yr. Three point sources in the vicinity of
the far-infrared (FIR) peak are also found to be early B type stars. The remaining
(predominantly low mass) members of the cluster are best represented
by a 1 - 3 Myr pre-main-sequence (PMS) population. $HK$ band spectra of
two bright and five faint point sources in the cluster confirm that
the results obtained from the photometry are good
representations of their young stellar object (YSO) nature. Thus, IRAS
19343+2026 is a young cluster with at least four early B-type stars
classified as young (10$^{4}$ - 10$^{5}$ yr), that are surrounded by a
somewhat older (1 - 3 Myr) population of low mass YSOs. 
Together, these results argue for a scenario in which low mass
stars form prior to massive stars in a cluster forming environment.
We compute the Initial Mass Function (IMF) for this cluster using the $K$-band
luminosity function; the slope of the IMF is shallower than
predicted by the Salpeter's mass function. The cluster mass, 
$M_{\rm total}$, is estimated to be in the range $\sim$ 307 \msun\ 
(from the data completeness limit) - 585 \msun\ (extrapolated down to 
the brown dwarf limit, assuming a certain IMF).

\end{abstract}


\section{Introduction}

One of the important issues in understanding massive star formation is
the sequence of formation of low and high mass stars in a cluster
forming environment \citep{zb08}. This issue has relevance when
explaining the nature of massive star formation in terms of either:
(a) low to intermediate mass stars accumulating matter while evolving toward
the main sequence, or (b) competitive accretion of massive cores
in a star forming cloud (\citealt*[]{mac02}; \citealt*[]{mac03};
\citealt*[]{zy07}).  Observations of embedded clusters generally
display mass segregated configurations, with a few young high mass
stars located at the cluster centers (\citealt*[]{mcn86}, and
references therein). Although an age estimation for the lower
mass pre-main-sequence (PMS) population is possible using
evolutionary tracks and colour-magnitude (CM) diagrams, the same is
not true for massive stars, since they are usually only a few in
number. Therefore, it is necessary to investigate clusters around high
mass protostellar object candidates, where the youth of the massive
stars is made clear by various other signposts, such as outflows,
compact HII regions, excess FIR emission, and so on. \citet{kum06}
searched for clustering around a sample of high mass protostellar
objects and found 54 embedded clusters among 215 candidates. The
detection of clustering in about 25\% of the sources (all targets
located away from the Galactic mid-plane) suggests that at least
one generation of low mass stars has already formed around these
massive protostellar candidates. However, the Two Micron All Sky
Survey (2MASS) data used by these authors did not effectively uncover
the lower mass population; the 2MASS data were not deep enough to
characterise the remaining cluster members. In contrast, a follow-up
of the same data sets by \citet{kum07} using the {\em Spitzer}-GLIMPSE
survey (\citealt*[]{church09}) did not find any clustering for the
sources in the Galactic mid-plane. This mid-infrared survey should
have been more successful at finding clusters because of the
better dust penetration.

In this paper, using high quality optical and infrared
observations of a high mass protostellar object in the Galactic mid-plane, 
we attempt to examine
such discrepancies, measure the stellar content, and understand the
sequence of low and high mass star formation in the associated
cluster. The high quality NIR photometric and spectroscopic data are
combined with {\em Spitzer} IRAC and MIPS observations to study the
target. Study of the massive star content is particularly aided by 
radiative transfer modelling of the SED. The target, IRAS 19343+2026,
is a high mass protostellar object candidate located at a kinematic
distance (measured using NH$_3$ line velocities) of 4.2 kpc, with
a FIR luminosity of 2.7$\times$10$^4$ L$_{\odot}$ \citep{mol96}.
This cluster is detected in the GLIMPSE survey (cluster 24 of
\citealt*{mcm05}).  Section 2 describes the detailed observations. The
photometric, spectroscopic and SED modelling results are also
presented in this section. The implications of the results are
discussed in Section 3. We then summarize our conclusions in Section 4.

\section{Observations and Data Reductions}

\subsection{$J, H, K$ imaging and photometry}

NIR photometric imaging observations were made at the 3.8~m United
Kingdom Infrared Telescope (UKIRT) with the facility imager UFTI
\citep{roche02}. UFTI houses a HAWAII-1 1024 $\times$ 1024 pixel array. 
The UFTI plate scale of 0.091$\arcsec$ gives an available field of
view (FOV) of $\sim 90\arcsec$. Photometric observations through
 $J$ ($\lambda=1.25$\mic, $\Delta\lambda=0.16$\mic ),
 $H$ ($\lambda=1.64$\mic, $\Delta\lambda=0.29$\mic ) and
 $K$ ($\lambda=2.20$\mic, $\Delta\lambda=0.34$\mic )
broad-band filters were obtained for the IRAS source during the night
of 26$^{th}$ June 2002. An integration time of 60 sec was used in each
of the $J$, $H$ and $K$ band filters; averaging jittered exposures
yielded a total exposure time of 540 seconds in each band.  The mean
seeing measured was 0.6$\arcsec$ in the $K$-band images. A nine point
(3$\times$3) jittered observing sequence was executed to obtain data
that provided final mosaics with a total FOV of
$\sim$115$\arcsec\times$115$\arcsec$. We note that the signal-to-noise
ratio at the edges of these mosaics is lower than within the central
90\arcsec~area. Standard data reduction techniques involving dark
subtraction and median-sky-flat-fielding of the jittered object
frames were applied. The $K$-band image of IRAS 19343+2026 is shown in
Fig.~1, with an overlay of {\em Spitzer} MIPS 24 $\mu$m contours.

Subsequent to our observations, this region was recently covered by
the UKIRT Infrared Deep Sky Survey (UKIDSS; \citealt*[]{lawrence07})
Galactic Plane Survey (GPS). The GPS is an ambitious survey of the
Northern Galactic plane \citep{lucas08}. The aim of the survey is 
to map 1800 square degrees of the plane ($|$b$|$ $<$ 5 deg) in 
$J$, $H$, and $K$ to a depth of $J$ $\sim$ 20.0, $H$ $\sim$ 19.1, 
$K$ $\sim$ 19.0 at 
sub-arcsecond resolution. UKIDSS employs the Wide Field Camera 
(WFCAM; \citealt*[]{cas07}) at UKIRT. WFCAM contains four Rockwell 
Hawaii-II (HgCdTe 2048x2048 pixel) arrays spaced by 94\% in the 
focal plane.  With a pixel scale of 0.4\arcsec\, the field of view 
of each array is 13.7\arcmin. All UKIDSS survey data are reduced by 
the Cambridge Astronomical Survey Unit (CASU) and are distributed via 
the WFCAM Science Archive 
(WSA\footnote{http://surveys.roe.ac.uk/wsa/index.html}).
The GPS data sensitivity
is comparable to our observations ($K$ =18 mag), albeit with a relatively
poor seeing (1\arcsec) and a coarse pixel scale of 0.2\arcsec.
Fig. 2 shows the 50$^{th}$ nearest-neighbour (NN) density map
(\citealt*{sch08}, and references therein) of the UKIDSS $K$-band
source counts. From this figure the cluster center was found to be
at $\alpha_{2000} = 19^{\rm h}36^{\rm m}31^{\rm s}, \delta_{2000} =
+20^{\rm \deg}33^{\rm '}03^{\rm ''}$.  For the purpose of evaluating
the foreground/background contamination, we have chosen a ``control
region'' $\sim$ 2$\arcmin$ North of the cluster center 
at $\alpha_{2000} = 19^{\rm h}36^{\rm m}32^{\rm s}, 
\delta_{2000} = +20^{\rm \deg}34^{\rm '}48^{\rm ''}$ 
(see Fig. 2).

We utilised tasks available in the Image Reduction and Analysis
Facility ({\sc iraf}) package for our photometric analysis. {\sc
DAOFIND} was used to identify sources in each image. A point spread
function (PSF) model was computed by choosing stars of different
brightness that were well spaced out in our images. Photometry was performed
using the {\sc DAOPHOT} package. Aperture corrections were determined by
performing multi-aperture photometry on the PSF stars. The
instrumental magnitudes were calibrated to the absolute scale using
observations of UKIRT faint standard stars; FS 29, FS 35 \& FS
140 \citep{haw01}. These standards were observed over a range in 
airmass (1.05 - 1.79) that was comparable to the target observations. 
The resulting photometric data are in the natural system of the Mauna
Kea Consortium Filters \citep{st02}.  For the purpose of plotting
these data, we converted magnitudes to the
\citet{bb88} (hereafter BB) system, since the main-sequence references
are in the BB system. To do this, we first converted the Mauna Kea
system to the CIT system and then to the BB system using equations
given by \citet{haw01}.

Representative sub-images consisting of stars and nebulosity were
chosen to determine the completeness limits.  Limits were established
by manually adding and then detecting artificial stars of differing
magnitudes. By determining the fraction of stars recovered in each
magnitude bin, we have deduced 90\% completeness limits of 19.3, 19.0
and 17.5 magnitudes in the $J$, $H$ and $K$ bands, respectively. Our
observations are absolutely complete (100\%) to the levels of 17.3,
17.2 and 16.2 magnitudes in $J$, $H$ and $K$,
respectively. Photometric analysis was carried out using data with
photometric errors of less than 0.1\,mag. Absolute position
calibration was achieved using the coordinates of a number of stars
from the 2MASS catalog. The astrometric accuracy of the data presented
in this work is better than 0.5\arcsec. The sources are saturated at
$K$ $<$ 12. For such bright sources, 2MASS Point Source Catalog data
were used. 

We also extracted source magnitudes from the UKIDSS images using
the methods described above, although our UFTI images (rather than
2MASS data) were used to calibrate the UKIDSS data. The UKIDSS
observations were used to create a density map of the region (see
Figure 2), and to obtain photometry over a larger area than was
observed with UFTI. 

\subsection{HK band spectroscopy} 

NIR $HK$ band spectra of several stars in the target field were
obtained using the UKIRT 1 - 5 $\mu$m Imager Spectrometer (UIST)
\citep{rh04} on the night of 21$^{st}$ July 2005. UIST has a 1024 $\times$
1024 InSb array and a 0.12 arcsec per pixel plate scale. The
observations were made using a 120 arcsec long slit with a width of 4
pixels. The $HK$ grism was used, which allowed complete wavelength
coverage from 1.395 $\mu$m to 2.506 $\mu$m with a spectral resolution
of R $\simeq$ 500. The exposure time per frame was set to 60 sec; 8
exposures were obtained per slit position, resulting in a total
integration time of 8 minutes. The 120\arcsec~long slit was aligned
carefully to cover multiple sources in a single position. This
configuration and integration time provided a 5$\sigma$ point source
sensitivity of $\sim$ 16 mag in the $HK$ bands. The target was nodded
up and down the slit in an ABBA fashion; B exposures were
subtracted from adjacent A exposures to give
sky-subtracted 2-dimensional spectral images.
The spectral images were averaged and the positive and negative
beams of stellar spectra were then optimally extracted. These
resulting ``group'' spectra were corrected for telluric absorption and
flux calibrated by division with a similarly-observed standard star
spectrum. The G2V source HIP102189 was used for this
purpose. (Note that the standard star spectrum was first divided by a
black body template of appropriate temperature to preserve the
intrinsic slopes of the calibrated target spectra.)

The initial reduction steps (i.e. dark and flat
correction) were performed using the UKIRT/Starlink data reduction
pipeline ORAC-DR \citep{cav08}; the extraction and analysis of the
spectra were done with IRAF and IDL.

\subsection{Optical Imaging and Photometry}

Bessell {\it BVRI} images of the region associated with the IRAS
19343+2026 region were obtained with the 2-m Himalayan {\it Chandra}
Telescope (HCT) of the Indian Astronomical Observatory (IAO), Hanle,
India, on 16$^{th}$ March 2005.
The Hanle Faint Object Spectrograph Camera (HFOSC), which is
equipped with a SITe 2K $\times$ 4K pixel$^2$ CCD, was used. 
With a pixel scale of 0.296$\arcsec$, the FOV of
HFOSC, where only the central 2K $\times$ 2K region is used for
imaging, is $\sim$ 10 $\times$ 10 arcmin$^2$. Further details on the
telescope and the instrument can be found at
\mbox{http://www.iiap.res.in/iao/hfosc.html}. Observations were
carried out under good photometric sky conditions. The typical
seeing [full width at half-maximum (FWHM)] during the period of
observations was $\sim$ 1.8$\arcsec$. Bias and flat frames were
obtained at the beginning and at the end of the observing
night. Photometric standard stars around the SA 111-775 region 
\citep{lan92} were observed to obtain the transformation coefficients.

The data reduction was again carried out using IRAF tasks. 
Object frames were flat-fielded using
median-combined normalized flat frames. Identification and
photometry of point sources were performed using the DAOFIND and
DAOPHOT tasks, respectively. Photometry was obtained using the
PSF algorithm ALLSTAR in the DAOPHOT package \citep{stet87}. The
residuals to the photometric solution were $\le$0.05 mag.

\subsection{Spectral energy distribution modelling}

The {\em Spitzer}-GLIMPSE survey IRAC and MIPS
images and IRAC photometry of this target were analysed using the
IRSA-IPAC image cutouts and GATOR facilities. MIPS 24 $\mu$m photometry
was extracted using the APEX single frame pipeline on the
Post-Basic-Calibrated Data. The photometric data in the {\em Spitzer}
bands (3.6 - 24.0 $\mu$m) were combined with 2MASS NIR photometry and
IRAS data to construct SEDs for the four brightest sources in the
region -- the massive young star candidates. The online SED fitting
tool developed by \citet{rob07} was used to fit the resulting SEDs in
the 1 - 24 $\mu$m bands. The SED fitting tool is based on
matching the observed SEDs with a large grid of precomputed
radiative transfer models. The models assume an accretion scenario
with a star, disk, envelope and bipolar cavity, all under radiative
equilibrium. While the mass and age of the star are uniformly
sampled within the grid limits, the radius and temperature are
interpolated using evolutionary models. 
The optical and NIR data points constrain the stellar
parameters, the mid-infrared fluxes constrain the disk parameters
and the far-infrared points are sensitive to envelope emission (see
\citealt*[]{rob06} for more details). The SED fitting tool has been
successfully tested on low mass stars and shown to produce reliable
estimates of physical parameters by comparing against values
obtained by other independent measurements. \citet{grave09} discuss
the implementation of this tool on a large sample of massive
protostellar objects. For fitting purposes a 10\% error was assumed
on all fluxes. The weighted mean (weights being the inverse of
$\chi^2$) of each physical parameter was computed for all models that
satisfied the criteria ${\chi}^2 -{{\chi}^2}_{best} < 3$, where
$\chi^2$ is the statistical goodness of fit parameter measured 
per data point. 

\section{Results} 

A rich cluster of stars can be seen in our $K$-band image (see
Fig.~1), the brightest stars in the center of the field coinciding
with the peak of the MIPS 24 $\mu$m contours. The two straight
lines in Fig. 1 mark the slit positions used to obtain the spectra of
the stars numbered 1 to 7. Stars 1, 3 and 8 were modelled using 1 -
24 $\mu$m SEDs. Star 9 was not detected in the MIPS 24 $\mu$m
band. The brightest stars in the field are expected to be massive
stars, while the fainter population represents the low mass
members. In Fig. 3 (see the online electronic version of this
article for a colour plot) a colour composite of the {\em Spitzer}
infrared images is shown. The three-colour composite image was made
using the {\em Spitzer} IRAC 3.6 $\mu$m, IRAC 8.0 $\mu$m and MIPS 24
$\mu$m images coded as blue, green and red, respectively. Notice the
bipolar shape of the nebula and the FIR excess emission appearing as
red.  The stars 8 \& 9 are associated with excess emission at 24
$\mu$m, as can be seen from the red colour surrounding these two
stars. Also, the rich cluster of stars seen in Fig. 1 is not visible
in this composite image. We will comment further on this issue
in Sect. 4. In the following, we first discuss the photometric analysis
of the whole sample, then the IMF of the cluster region, followed by 
spectra of the seven representative
sources (stars 1-7), and finally the SED fitting analysis of the
brighter massive star candidates.

\begin{figure}
\includegraphics[width=\columnwidth]{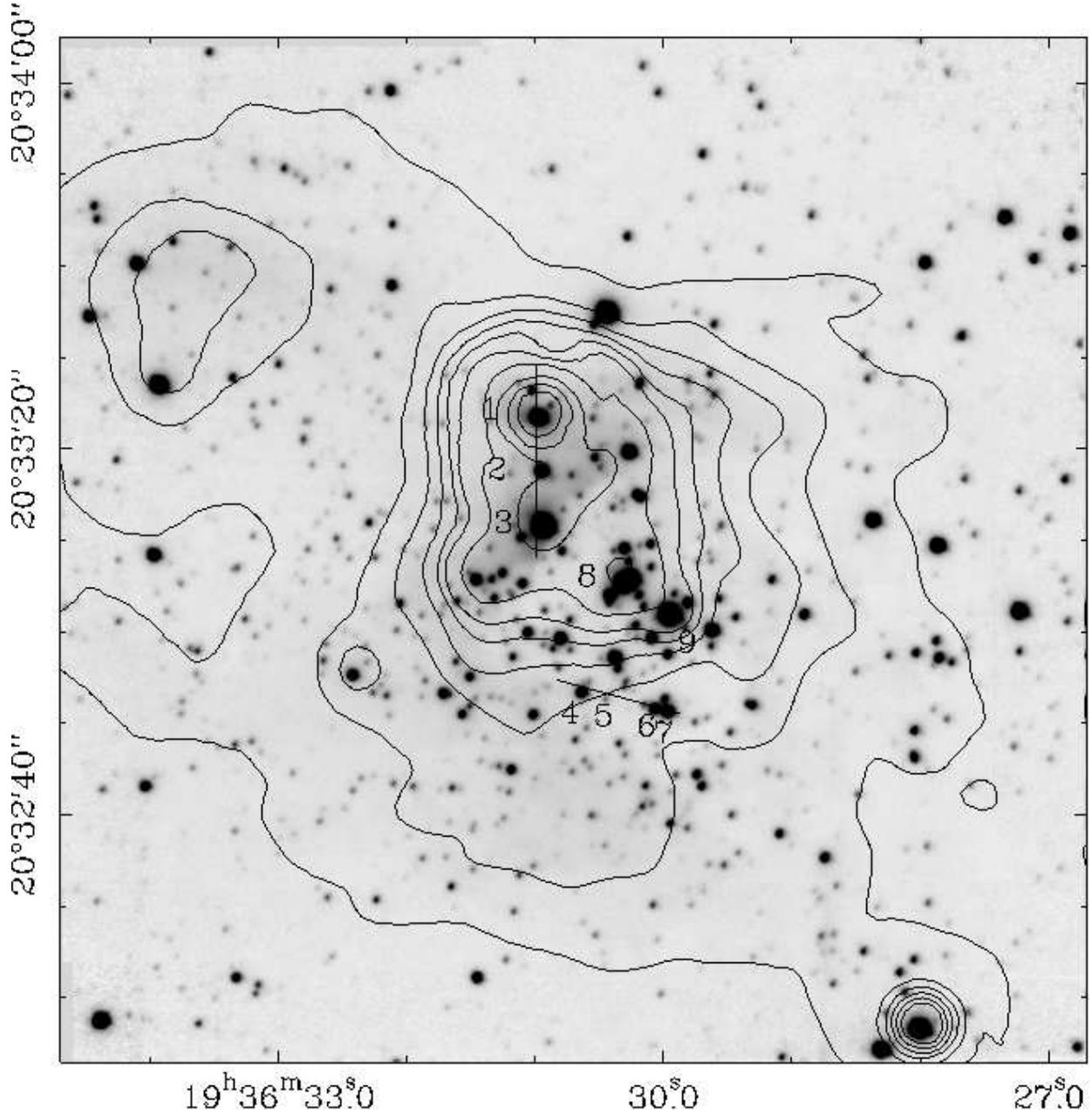}
\caption{A $K$-band image of the IRAS\,19343+2026 region shown with a
logarithmic greyscale stretch. Contours represent the MIPS 24 $\mu$m
emission. The solid lines mark the positions of the slits used to
obtain $HK$ band spectra of the stars numbered 1--7; note that spectra were
not obtained for the bright NIR sources labelled 8 and 9. 
The abscissa (RA) and the ordinate (DEC) are in J2000.0 epoch.}
\label{fig1}
\end{figure}

\begin{figure}
\includegraphics[width=\columnwidth]{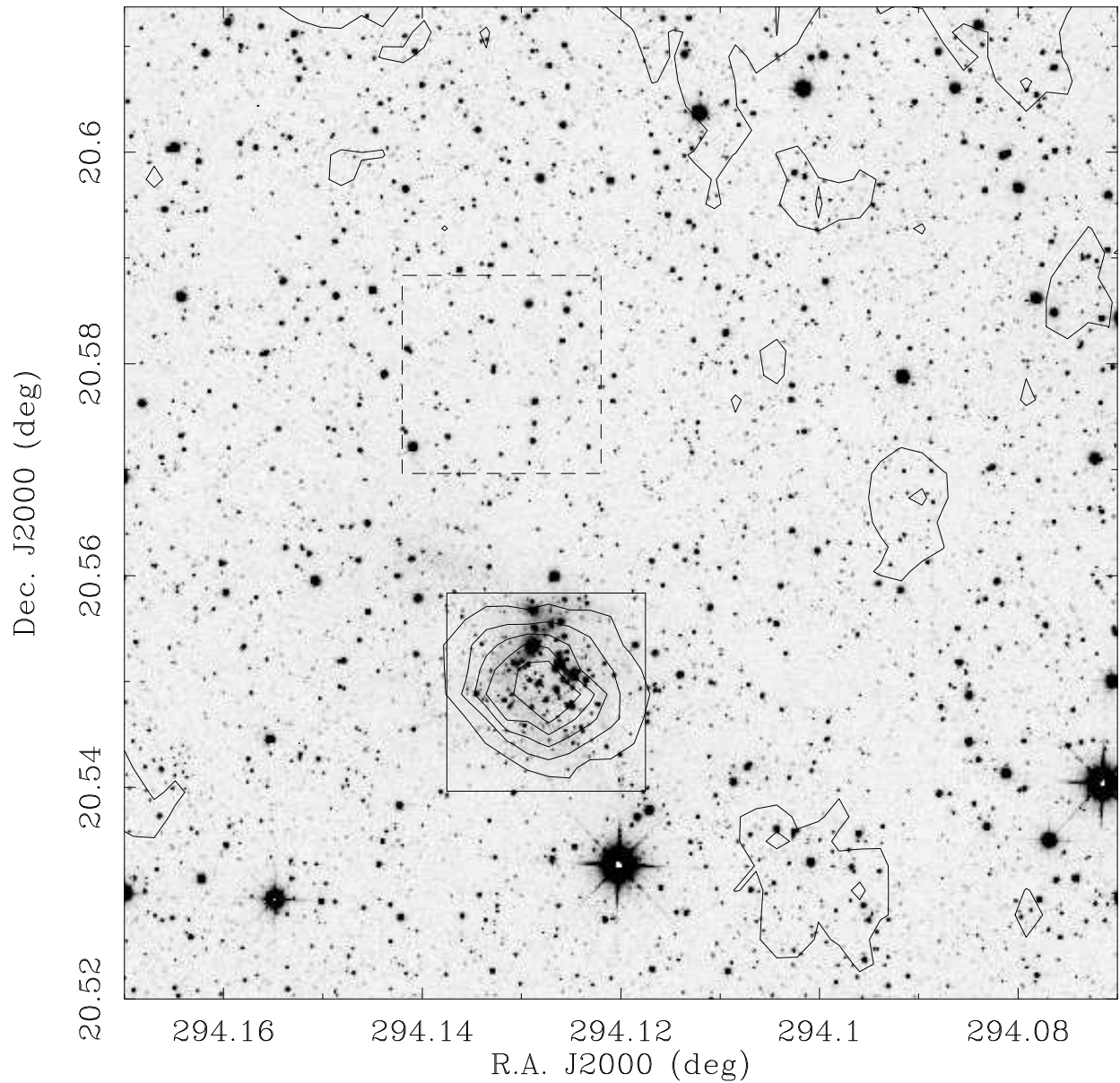}
\caption{50$^{th}$ nearest-neighbour density map of UKIDSS
$K$-band source counts around the IRAS\,19343+2026 region. 
The contours are plotted over the UKIDSS image; the contour
level starts at 111 stars per arcmin$^2$ and increases in steps of
28 stars per arcmin$^2$. The solid and dashed boxes represent the chosen
cluster and control field regions.} 
\label{fig2}
\end{figure}

\begin{figure}
\includegraphics[width=\columnwidth]{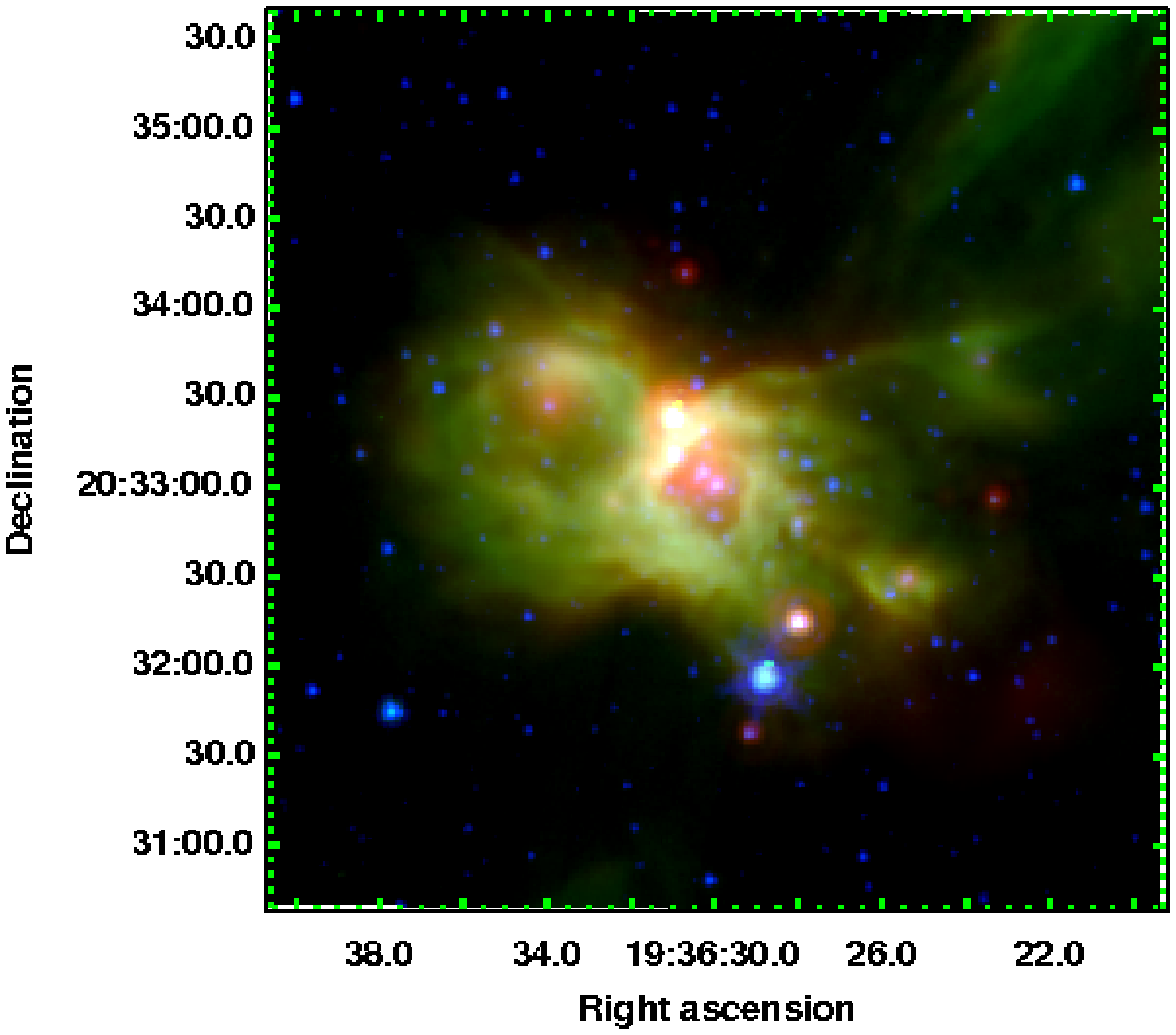}
\caption{Three-colour composite made from {\em Spitzer} images
(for a colour representation see the online electronic version of this
paper). The MIPS 24 $\mu$m, IRAC Ch4 (8.0 $\mu$m) and IRAC Ch1 (3.6
$\mu$m) images are coded red, green and blue respectively. The
abscissa and the ordinate are in J2000.0 epoch.}
\label{fig3}
\end{figure}

\subsection{Photometric Analysis} 

The $JHK$ photometry of the point sources in the IRAS 19343+2026
region was used to construct colour-colour (CC) and CM diagrams. A
combination of CC and CM diagrams made using various optical and
infrared bands were then used to evaluate the membership of the cluster
against a control field, and to evaluate the cluster properties. The
total number of point sources (with magnitude errors $<$ 0.1 mag),
detected in the region shown in Fig. 1 was 333, 688 and 875 in the $J$,
$H$ and $K$ bands, respectively. 307 stars are common to all three
bands; 607 stars appear in the $H$ and $K$ bands. The UFTI $JHK$
photometric data, along with the positions of the stars, are given in
Table A.1; the complete table is available in electronic form as part
of the online material.

Photometry of the images obtained from 
the UKIDSS-GPS survey was performed over a larger area 
($\sim 7\arcmin \times 7\arcmin$). The analysis of both the cluster 
and control regions was carried out in a similar way. In order to
evaluate the mass function (see Sect. 3.2) of the cluster region,
we have chosen a box size of 1.2\arcmin\,
centered on the cluster and control regions, as mentioned earlier (see
Sect. 2.1).

Fig.\,4 shows the $J-H/H-K$ CC diagram for the cluster field. The
solid ({\it bottom-left}) and broken heavy curves represent the 
main-sequence dwarf and
giant stars, respectively. The dotted line indicates the locus of T
Tauri stars \citep{meyer97}, while the dashed-dotted line represents
the HAeBe locus \citep{ladadams92}. The three dashed parallel lines are the
reddening vectors. We adopt a slope consistent with $E(J-H)/E(H-K)$ =
1.9 (BB system), which is appropriate to an interstellar reddening law
of R = 3.12 \citep{whittet80}. The CC diagram was used to identify the
reddened population that falls in the region occupied by T Tauri and
HAeBe stars. Stars that lie outside the region of reddened
main-sequence objects (i.e. redward of the reddening line drawn from
the base of the main-sequence dwarf branch, that is, redward of the
middle of the three vectors) are YSOs with intrinsic colour
excesses. We shall refer to these as ``probable cluster members''. By
de-reddening the stars on the CC diagram that fall within the first
two reddening
vectors (encompassing the main-sequence and giant stars) to the dwarf
locus, a visual extinction ($A{\rm_V}$) for each star was
calculated. The individual extinction values range from 0 to 20 mag,
resulting in an average extinction of $A{\rm_V}$ = 7.6 mag.

\begin{figure}
\includegraphics[width=\columnwidth]{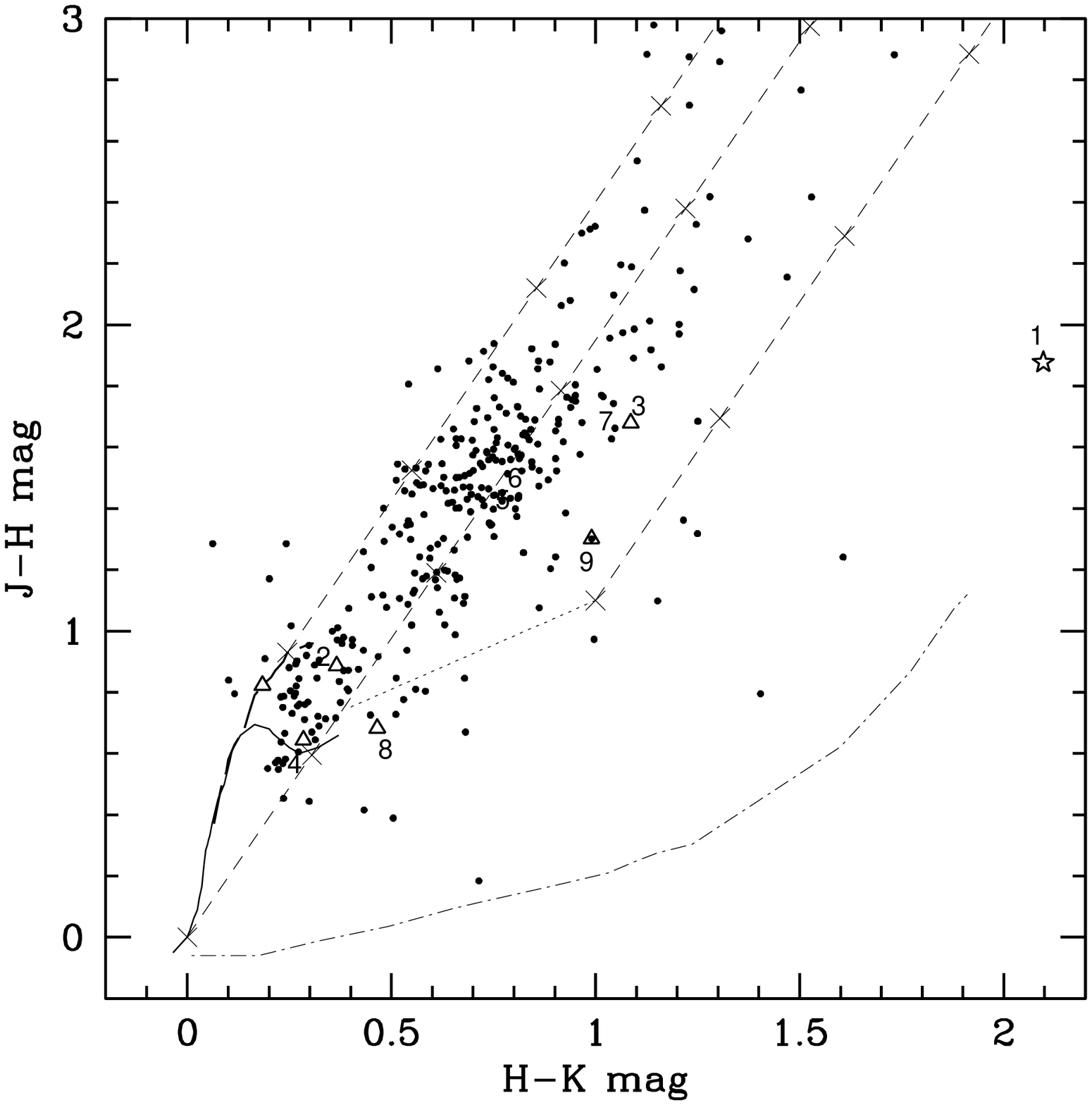}
\caption{CC diagram of the region around IRAS 19343+2026 
constructed from the UFTI data. The solid ({\it bottom-left}) 
and broken heavy curves
represent the main-sequence dwarf and giant stars, respectively, taken
from \citet{bb88}. The dotted line indicates the locus of T Tauri
stars \citep{meyer97}; the dashed-dotted curve represents the HAeBe
locus \citep{ladadams92}. The dashed parallel lines are reddening
vectors on which crosses separated by $A_{\rm V}$ = 5\,mag have been
marked. The triangles mark the brightest NIR sources in the region, 
while the numbers refer to the sources labelled in Fig. 1. 
The star symbol (star 1) marks the 2 $\mu$m source 
that coincides with the FIR peak.}
\label{fig4}
\end{figure}

Fig.\,5 shows the UKIDSS GPS $K/H-K$ CM diagram for the
cluster and the control region.  From the control region plot ({\it
middle panel}) it is possible to clearly identify the dwarf and giant
branches of the stellar population, i.e. sources with an $H-K$
colour of less than 1.0. In contrast, the cluster region ({\it left
panel}) displays the dwarf branch (the group of stars running
vertically at $H-K$ $\sim$ 0.4) and a significant number of stars with
$H-K$ colours around 1. These stars are essentially cluster members which are
also identified on the CC diagram as probable cluster members. 
The control field also has very few stars with $H-K$ $<$ 1 as
compared to the cluster region. Therefore, we find that the contamination 
from foreground/background stars in our cluster region is minimal
compared to the statistics of the probable cluster members. 
However, $H-K$ colour alone is no good for distinguishing between a
reddened background star and a young star. Therefore, we have used
a statistical approach to remove contaminating field stars from 
the $K/H-K$ CM diagram of the cluster region (we will discuss this
further in detail in the following Sect. 3.2). The cluster member stars 
are clearly seen in the statistically cleaned CM diagram ({\it right panel}). 

\begin{figure*}
\includegraphics[width=17cm]{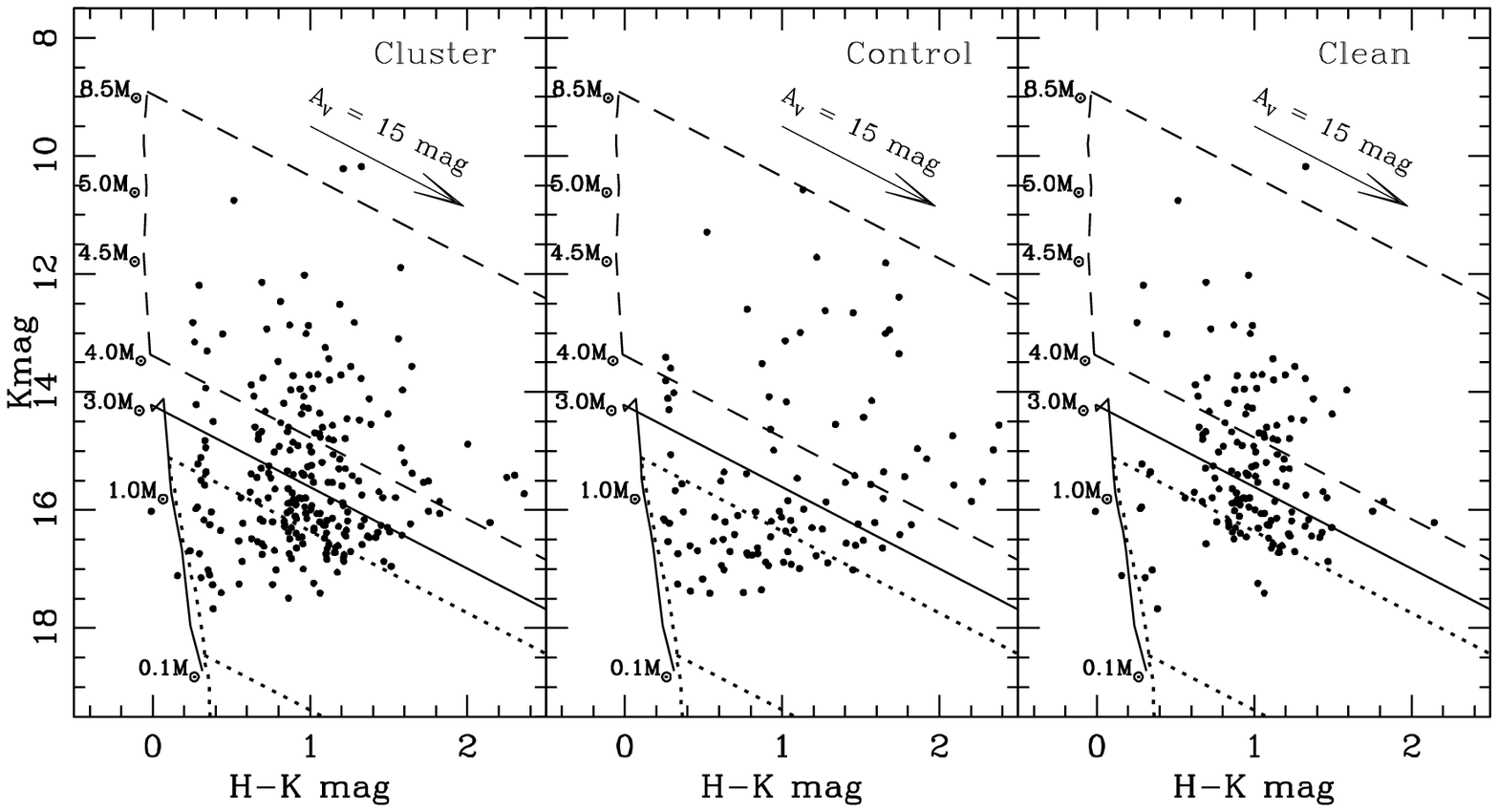}
\vspace*{-7cm}
\caption{$K/H-K$ CM diagrams of the IRAS 19343+2026 cluster region ({\it left}), 
the control region ({\it middle}) and the statistically cleaned cluster 
region ({\it right}) from the UKIDSS GPS data. The solid and dotted curves
show the PMS tracks from \citet{palla99} and \citet{bar98},
respectively, for an assumed age of 3 Myr for low mass range ($M$ $<$ 3 \msun),
while the dashed curve denotes the PMS track for 3 Myr age \citep{ls01}
for intermediate mass range
(4 $<$ $M$/\msun~$\leq$ 8.5). The parallel slanting lines identify the
reddening vectors.}
\label{fig5}
\end{figure*}

Our aim here is to estimate the approximate age of the lower mass
population by comparing our data with theoretical evolutionary tracks.
The low mass evolutionary tracks have been tested on various
observational data of young stellar clusters and are thought to be
good representations. The evolutionary models for high mass stars
are more complicated and varied, however, and are not well tested
against observational data sets. Therefore, we first use the
pre-main-sequence (PMS) isochrone fitting for the lower mass
population. Further, it should be noted that a CM diagram
that uses only one colour may not be very effective in describing
the less understood massive young stars.

Since the stellar photosphere's are better represented by shorter
wavelength data such as the optical, $J$ and $H$ bands, we use
these data when comparing source photometry with the model
isochrones. In the region of Fig. 1, all optically detected stars
in the $BVRI$ bands were plotted on a $V$ vs $V-I$ CM diagram. All
but seven of the optically detected stars are foreground
stars. Since seven sources is not statistically significant for
the young cluster, we instead plot the $J$ vs $J-H$ CM diagram in
Fig. 6. In this plot the slanting arrow denotes the reddening
vector associated with 15 mag of extinction. YSOs identified in
the $JHK$ CC diagram in Fig.\,4 are shown as open circles;
triangles mark the bright stars.

PMS evolutionary tracks \citep{palla99} for ages 1, 3 and 5 Myrs are
shown in Fig. 6 with dotted, short-dashed, and long-dashed lines,
respectively. We have assumed a distance of 4.2 kpc \citep{mol96} and
have reddened the isochrones with the average extinction of 
$A_{\rm V}$ = 7.6 mag estimated from the $JHK$ CC diagram analysis in
Sect. 3.1. Given the poor statistics for this distant cluster, 
it is not easy to derive the age by isochrone fitting. Even for
regions with good statistics, it is quite difficult to constrain the
age because most low mass stars spend the bulk of their PMS time
on the Hayashi track, which is the {\em vertical} part of each
isochrone. However, a small fraction of observed young stars will
coincide with the brief evolutionary phase associated with the
Henyey track, which is the {\em horizontal} transition between the
Hayashi track and the Zero-Age-Main Sequence (ZAMS) (see
e.g. \citealt*{ascenso07}).  

In Fig. 6 we note that there is a large population of 
low mass/low $J$ magnitude candidate young stars (open circles) 
that lie to the {\em right} of the 1 Myr isochrone (dotted
line). Some of these sources may be Class I protostars associated 
with the cluster. However, most will be Class II sources (T Tauri 
stars) with an age of about 1 Myr; many of these sources lie to
the right of the 1 Myr isochrone because of extinction which, to
these embedded sources, will be higher than the 
$A_{\rm V}$ $\sim$ 8 mags used to deredden the PMS isochrones. 
However, there is 
also a population of candidate low mass young stars that lie to 
the {\em left} of the 1 Myr isochrone. These sources represent a 
more evolved group of young stars that are probably 3 Myr or older. 
Indeed, there are five probable cluster members that align 
horizontally and may therefore represent the Henyey part of the 
3 Myr isochrone (short dash). Hence, we estimate that there is a 
low mass population of stars associated with IRAS 19343+2026 that 
is best represented by an age of 1 Myr or more.
The mass range plotted in the figure is from 0.6~\msun~to 3.0~\msun~ 
for the 3 Myr isochrone.
The 5 Myr isochrone (long dash; mass range 0.6 -- 2.5 \msun) may well 
be within the age spread of the cluster. However, it extends beyond 
the limits of the observed data points.

\begin{figure}
\includegraphics[width=\columnwidth]{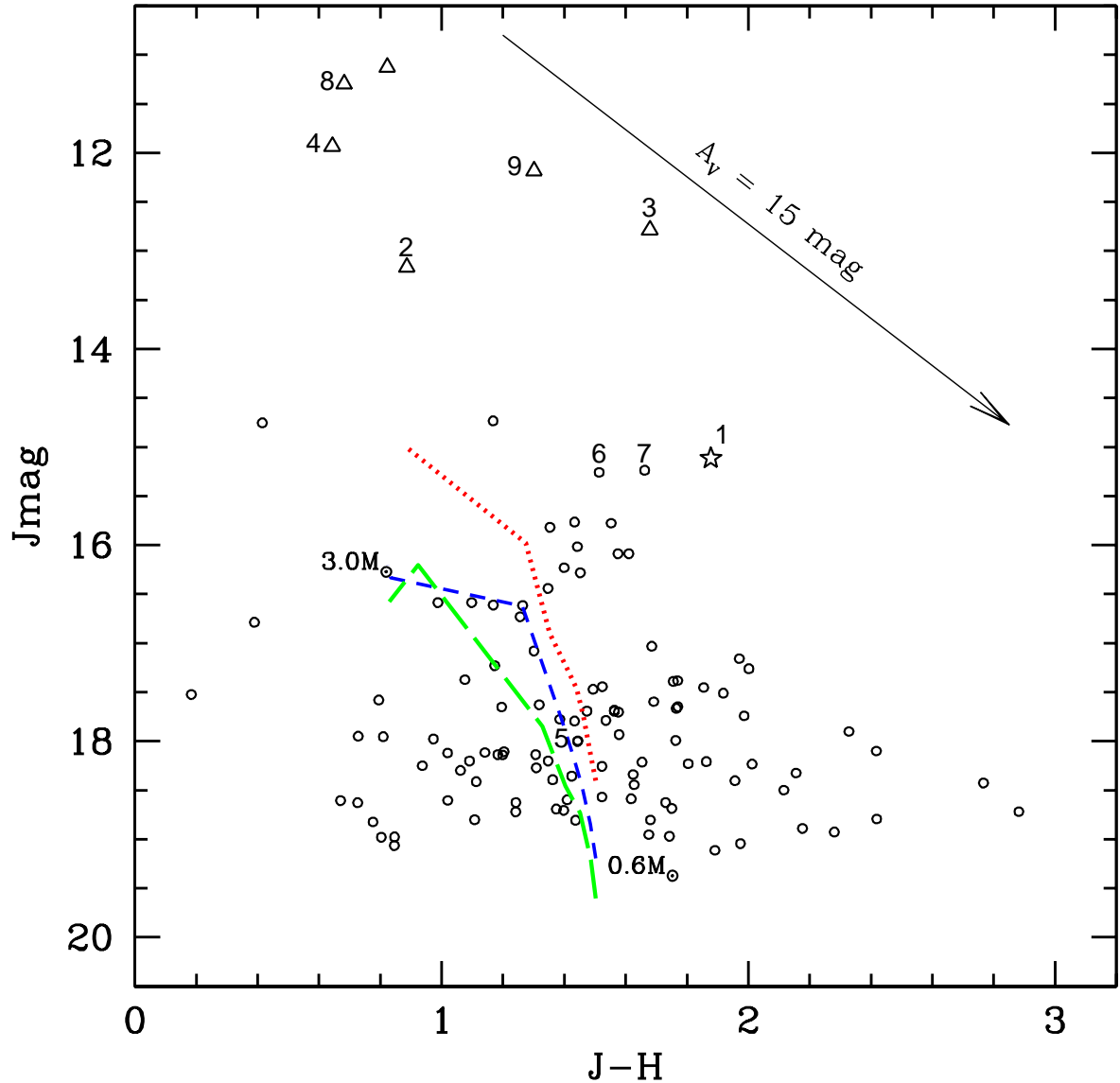}
\caption{$J$ vs $J-H$ CM diagram from the UFTI data. Open circles
are probable cluster members as estimated from the $J-H/H-K$ CC
diagram in Fig.~\ref{fig4}. The triangles mark the brightest 
NIR sources in the region, while the numbers refer to the sources 
labelled in Fig. 1. The star symbol (star 1) marks the 2 $\mu$m source 
that coincides with the FIR peak. The
PMS 1 Myr, 3 Myr (range = 0.6 \msun -- 3.0 \msun; marked) and 5 Myr
(range = 0.6 \msun -- 2.5 \msun) isochrones \citep{palla99} are 
drawn using dotted, short-dashed and long-dashed lines. The
slanting solid line shows the reddening vector.}
\label{fig6}
\end{figure}

One should also note that the $J$ vs $J-H$ diagram contains only
333 sources that are common to the $J$ and $H$ bands. This is a
considerable under-representation of the full sample, given that 875
sources are detected in the $K$ band and that many of the
fainter, redder sources will be detected only at longer
wavelengths. Nevertheless, the sample is statistically significant
when comparing with evolutionary tracks and obtaining an age estimate
for the low mass sources in this region.

\subsection{The Initial Mass Function}

As a next step to better understanding this region, we evaluate
the mass function of the cluster. In the NIR, young stars are
best detected in the $K$-band. Consequently, we use the
$K$-band Luminosity function (KLF) to evaluate this statistically
important parameter. Ideally, we want to build the KLF using UFTI data
which has a higher spatial resolution compared to the UKIDSS data. But
the estimate of the mass function should be compared with the mass
function of the control field, which is not possible using UFTI data,
since we do not have control field images obtained with
UFTI. The number of sources detected in the UFTI and UKIDSS images is
very similar (to within 5\%) for the same FOV. 
Therefore, we will
evaluate the initial mass function (IMF) using the UKIDSS data which
is uniformly calibrated for both the cluster and control regions.

By using the control region comparison, the differences arising due to
contamination from the foreground and background members are accounted
for. Although we identified probable cluster members (T Tauri and
related sources) in Sec. 3.1, this identification was based on using
only those stars that appear in the $J$, $H$ and $K$ bands. However,
there are roughly twice as many stars detected in the $K$-band, 
whose cluster membership can not be easily verified. Therefore, we
have used statistical criteria to estimate the number of probable
member stars in the cluster region. To remove contamination of field
stars from the cluster region sample, we have statistically subtracted the
contribution of field stars from the CM diagram of the cluster region
using the following procedure. For a randomly selected star in the
$K$, $(H-K)$ CM diagram of the control region (see Fig. 5), the
nearest star in the cluster's $K$, $(H-K)$ CM diagram within $K$ $\pm$
1.0 and $(H-K)$ $\pm$ 0.5 of the field star was removed. The
statistically cleaned $K$, $(H-K)$ CM diagram of the cluster region is
shown in Fig. 5 ({\it right}) which clearly shows the presence of PMS
stars in the cluster. With the help of this statistically cleaned CM
diagram, we compute the IMF \citep[e.g.,][]{ojha09}. The $K$-band
luminosities are corrected by the mean extinction determined by
dereddening the stars on the CC diagram (see Sect. 3.1) which is
$A_{\rm V}$ $\sim$ 8.0 mag. The de-reddened magnitudes were then used to construct
a reddening-corrected KLF. Stellar masses for the assumed 3 Myr old
cluster members were obtained using the evolutionary models of
Geneva \citep{ls01} and \citet{palla99}. The MF of the 
statistically cleaned cluster region sources was obtained by 
counting the number of stars in various mass bins and is
shown in Fig. 7. 
The vertical dashed and dot-dashed lines represent 90\% and
100\% completeness limits obtained for the UKIDSS GPS data, 
respectively.

The MF of the IRAS 19343+2026 region appears to rise monotonically
up to $\sim$ 1.0 \msun.  The MF has a slope ($dLog(N)/Log(M)$) of
-1.12$\pm$0.16 for the probable cluster members sample over the mass 
range 1.0 $<$ $M$/\msun~$<$ 6.3. The classical value derived by 
\cite{sal55} is -1.35. Comparison of the MF of the IRAS 19343+2026
region with that of the Trapezium cluster (slope = -1.21 for
$M$/\msun~$>$ 0.6) measured by \citet{muench02} reveals a resemblance,
in the sense that both MFs rise monotonically beyond $M$/\msun~$>$
0.6. The cluster region slope is shallower than the Salpeter value (-1.35).
This indicates that the star formation process is not yet
complete in the region. Star formation is ongoing, although previous
episodes of star formation activity have resulted in a significantly 
low mass content. This situation represents active star formation over
a period of 3 -- 5 Myr as observed in other examples of massive star
forming regions \citep{ascenso07}.  

We have also estimated the mass of the cluster by integrating the
observed mass function above the completeness limit ($log M_{\rm lim}$
$\sim$ 0.1 \msun~from Fig. 7).  A mass of $\sim$ 307 \msun\ is
obtained. For the masses below the completeness limit, we have
extrapolated the theoretical IMF slopes from \citet{kru01} down to the
brown dwarf limit (see Fig. 7). The integration over the IMF yields a
total cluster mass, $M_{\rm total}$ $\sim$ 585 \msun. This is
comparable to the masses of clusters such as those associated with the
Trapezium and Mon R2 \citep{ll03}. 

\begin{figure}
\includegraphics[width=\columnwidth]{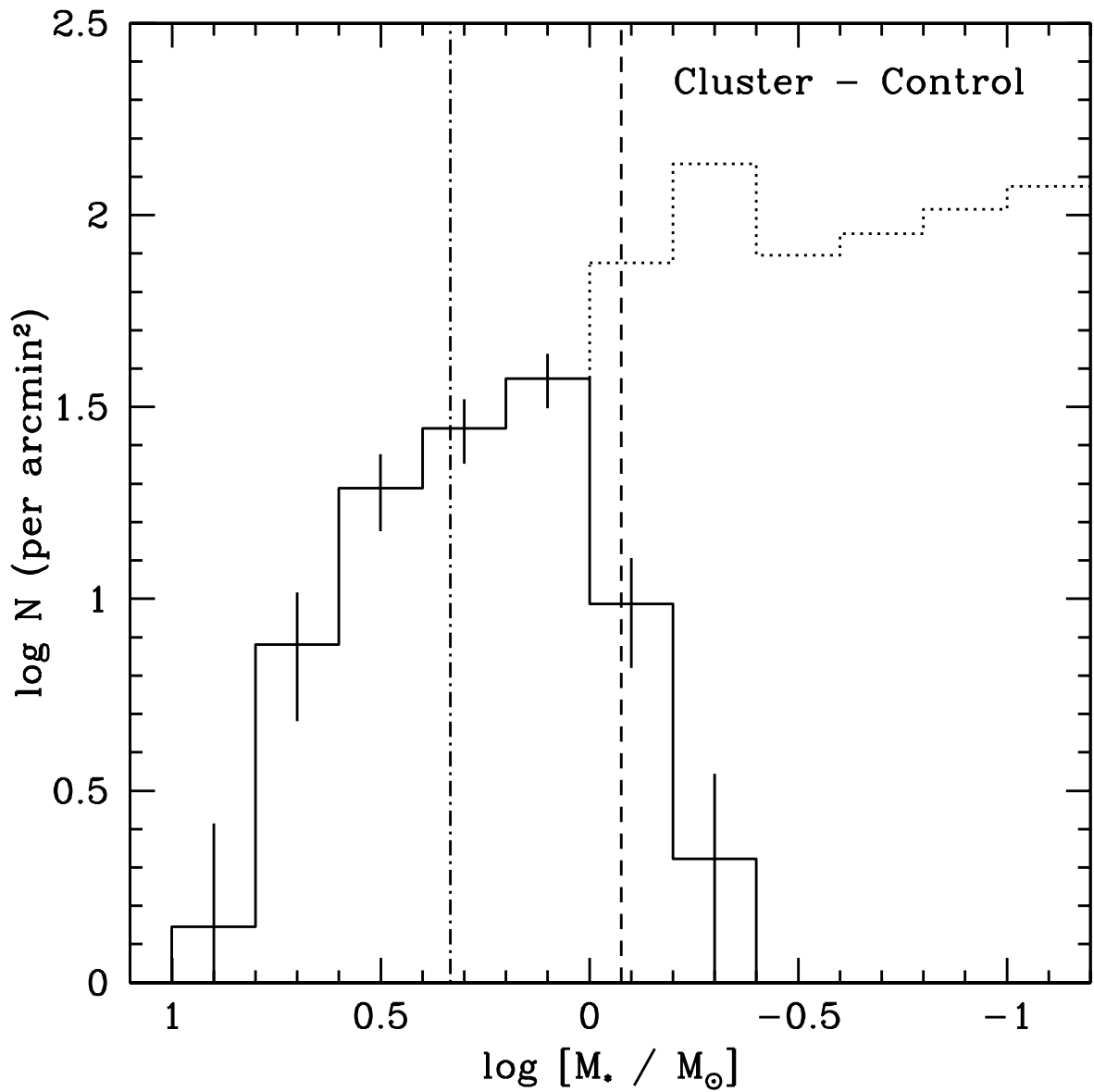}
\caption{The mass function (with $\pm\sqrt{N}$ error bars) of the 
statistically cleaned IRAS 19343+2026 cluster region derived from the UKIDSS data. 
The plot is based on the evolutionary models of \citet{ls01} and
\citet{palla99}, for an assumed age of 3 Myr. The vertical dashed and
dot-dashed lines represent 90\% and 100\% completeness limits,
respectively. The dotted line shows the theoretical IMF from
\citet{kru01}, for the masses below the completeness limit ($log
M_{\rm lim}$ $\sim$ 0.1 \msun).}
\label{fig7}
\end{figure}

\subsection{Spectroscopy}

Long-slit $HK$ band spectra of two bright and five faint sources were
obtained to investigate the nature of the stellar sources in the
cluster.  In Fig.\,1, the straight lines mark the positioning of the
slit which, for each exposure, covered multiple stars. The
numbers by the side of the line identify the stars seen through the
slit. In Fig.\,8 the spectra of each of these stars are shown. 
The stars 1, 2 and 3 are the bright stars
that are close to the FIR emission peak. Stars 4, 5, 6 and 7 are
fainter (presumably low mass) stars that lie in the central
regions of the cluster. Stars 3, 5, 6 and 7 
occupy the T Tauri zone in our $JHK$ CC diagram in Fig. 4,
while star 1 lies in the HAeBe zone of this plot.

The two bright stars (1 \& 3) that coincide with the FIR peak (see
Fig. 1) display the HI Bracket series of lines indicating
ionised emission.  Note that star 2, which lies in between stars 1 and
3 on the slit, does not show the same recombination lines. The HI
lines are thought to arise very close to the star, since the
120\arcsec~long slit encompasses the dense nebula and we see no
signature of extended nebular emission. As will be shown in the next
section, the SED modelling of these bright objects classifies
them as massive young stars. Star 1, the best massive
protostellar candidate in the region, shows intense Pa$\alpha$
emission, in addition to the HI recombination lines. Both stars, 1 and
3, display a rising slope long-ward of 2 $\mu$m.  Together, these
features indicate that stars 1 and 3 are early B type stars capable of
producing an ionised sphere around themselves \citep{oster89}.
Main-sequence B stars are characterised by HI absorption features
which become weaker for more massive stars. However, for O type
stars, helium absorption features appear and become prominent with
increasing mass \citep{han05}. Neither HI nor He absorption
features are visible in the bright stars 1 and 3, suggesting that the
photospheres are likely obscured by the surrounding compact HII region
and dense material. 

In contrast, the remaining spectra of fainter stars (4, 5, 6 \& 7) are
featureless, and display a slope that is either flat or rising
short-ward of 2 $\mu$m.  The absence of any emission features suggests
a relatively evolved T Tauri star (Antoniucci et al. 2008), while the
absence of any absorption features would suggest heavy veiling
\citep[and references therein]{gc96}. Therefore, the fainter stars 4,
5, 6 \& 7 are best matched by T Tauri type stars. The stars 6 \& 7 are
separated by a projected angle of 1.5\arcsec; both show a common
spectral feature at 2.4 $\mu$m. This broad feature was not
identified with any known spectral lines. The feature is likely to 
be a poorly removed telluric absorption feature (owing to poor sky 
beyond 2.3 $\mu$m) and/or a local artifact in the slit. We note that 
there is no such spectral feature known in previous literature, even in 
the most embedded sources with lots of emission lines.

\begin{figure}
\includegraphics[width=\columnwidth]{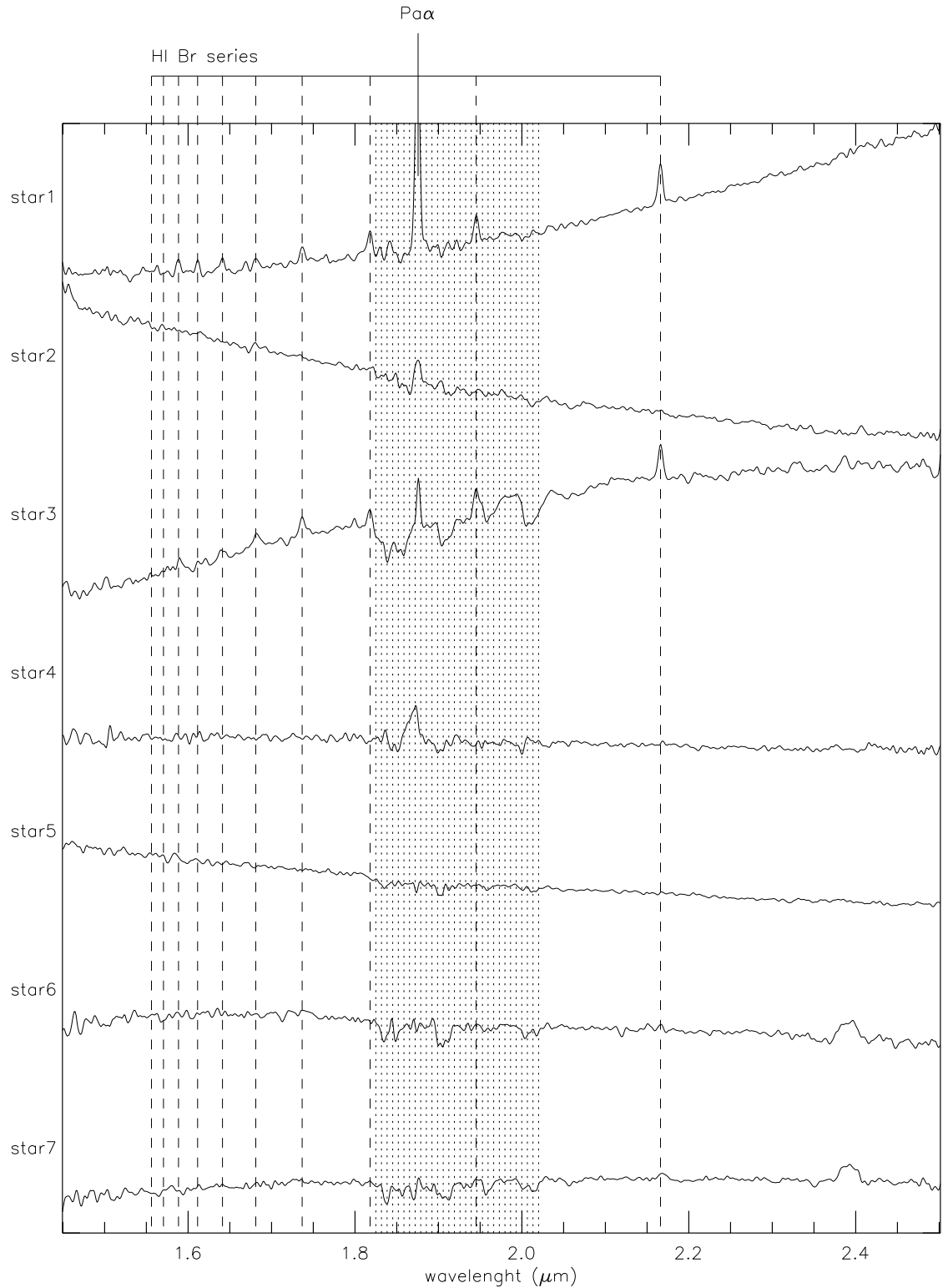}
\caption{$HK$ band spectra of the stars marked in Fig. 1. The
shaded area represents the region of poor atmospheric
transmission.}
\label{fig8}
\end{figure}

\subsection{SED modelling of bright stars}

The four bright stars that dominate the central region of the infrared
nebula, namely stars 1, 3, 8 and 9 (see Fig. 1), are resolved in the
{\em Spitzer} IRAC bands for which point source photometry is
available. Note that these four stars are the brightest NIR
sources associated with the IRAS 19343+2026 cluster (see Table 1). 
Only stars 1, 3 and 8 were detected in the MIPS 24 $\mu$m
images. Stars 3, 8 and 9 are also detected in our optical $BVRI$
images. Of these, star 8 is a multiple star with unreliable photometry
since the PSF is larger than the separation of the multiple
components. Therefore, the optical magnitudes of star 8 are used as
upper limits in the modelling. 

The SEDs of these four sources were modelled using the online SED
fitting tool developed by \citet{rob07}. The results of the SED
modelling are summarised in Table 2. In Fig.~9, the observed data
points (shown as black dots and filled triangles) 
and the fitted SED models are plotted.
The solid black line shows the best fit model while the grey lines
represent models that satisfy the criteria of ${\chi}^2 -
{{\chi}^2}_{best} < 3$, where $\chi^2$ is per data point. The
weighted means and standard deviations in Table 2 are calculated using
all of the parameters from the models that satisfy the above criteria.
The weight is the inverse of the $\chi^2$ value.  The standard
deviations quoted in Table 2 help the reader to visualize the spread
in the range of values between multiple models that satisfy the above
criteria.  The dotted line represents the photospheric emission
assumed in the model.

It is interesting to note that the FIR peak (star 1) is a young
massive star of $\sim$ 7.6 \msun~which is deeply embedded in an
envelope. The remaining bright stars are 6 - 7 \msun~objects with
nearly revealed photospheres surrounded by remnant dust envelopes
(indicated by the second peak in the SED models). These results
are consistent with the fact that star 1 is not optically visible,
unlike stars 3, 8 and 9. The VLA NVSS survey shows centimeter
free-free continuum emission coincident with the central region of
this nebula. The cm continuuum emission may represent the ionised gas
in the central nebula associated with these four early B type stars.

\begin{figure*}
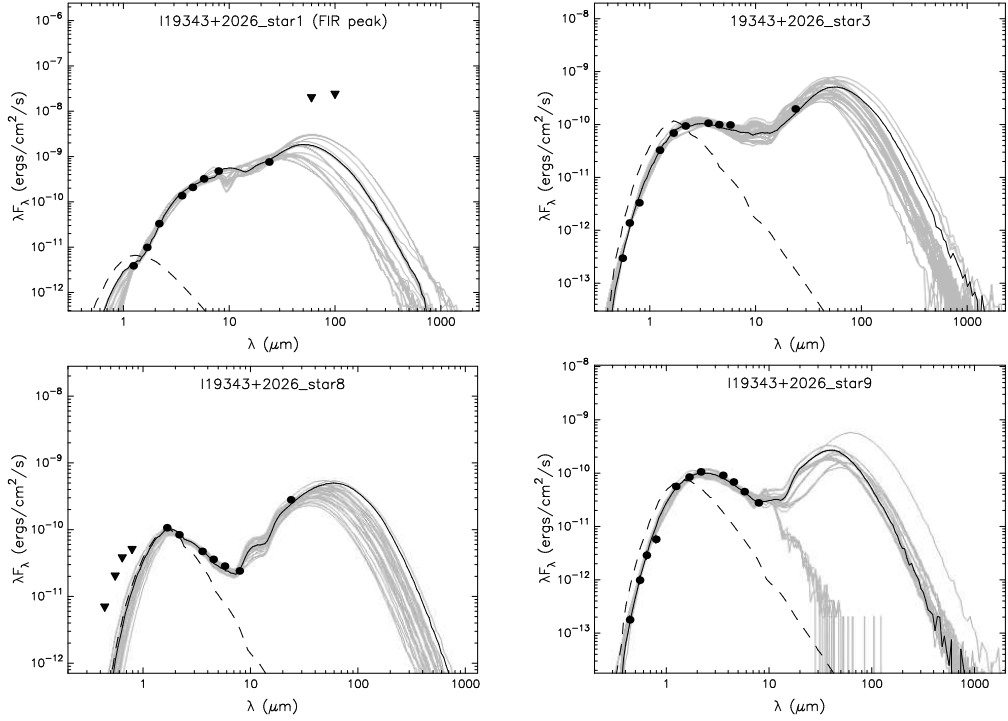

\includegraphics[width=6.3cm]{fig9a.eps}
\qquad
\includegraphics[width=6.3cm]{fig9b.eps}
\includegraphics[width=6.3cm]{fig9c.eps}
\qquad
\includegraphics[width=6.3cm]{fig9d.eps}
\caption{SEDs for the FIR peak, star 1, and three other
candidate massive young stars. The black dots display photometric
data points from 1 - 24 $\mu$m from 2MASS and {\em Spitzer}; filled
triangles mark optical or IRAS 60 $\mu$m and 100 $\mu$m data. 
The solid grey
curves denote the family of fitted models; the black curve
represents the best fit (the dashed curve indicates the photosphere
emission input to produce the best fit model). Note that these
bright sources were saturated in the UFTI and UKIDSS data.}
\label{fig9}
\end{figure*}

\begin{table*}
\caption{Photometry used for SED modelling$^{a,b}$}
\begin{tabular}{ccccccccccccc}
\hline
RA (2000) & DEC (2000) & $B$ & $V$ & $R$ & $I$ & $J$ & $H$ & $K$ & 3.6$\mu$m & 4.5$\mu$m & 5.8$\mu$ & 8.0$\mu$m \\
deg & deg& mag & mag & mag & mag & mag & mag & mag & mag & mag & mag & mag\\
\hline
294.12900 & 20.55662 & - & - & - & - & 14.99 & 13.18 & 11.11 &  8.10 & 6.90 &  5.68 & 4.27 \\            
294.12899 & 20.55319 & - & 19.60 &  17.50 & 16.20 & 12.68 & 11.07 &  9.98 & 8.37 & 7.70 & 6.96 & - \\           
294.12634 & 20.55140 & 16.50 &  15.00 & 13.88 &  13.22 & 11.22 & 10.60 & 10.11 & 9.25 & 8.81 & 8.31 & 7.51 \\           
294.12489 & 20.55042 & 20.50  & 18.30 & 16.70  & 15.60 & 12.09 &  10.85 & 9.86 & 8.54 & 8.11 & 7.82 & 7.36 \\
\hline
\end{tabular}

$^a$0.1 mag (10\%) error is assumed on all magnitudes. See text for details.

$^b$The UFTI $J, H, K$ photometry of all the cluster members is available in 
the online table.

\end{table*}

\begin{table*}
\centering      
\caption{SED modelling results} 
\label{photstat} 
\begin{tabular}{cccccccc}    
\hline  
Source & ${\chi}^2$ & Mass (\msun) & Log(Age) yr & log(M$_{disk}$) \msun & log(M$_{env}$) \msun & log($\dot{M}_{disk}$) \msun yr$^{-1}$ & log($\dot{M}_{env}$) \msun yr$^{-1}$\\
\hline
Star 1 & 10.0 & 7.6$\pm$0.6 & 5.2$\pm$0.6 & -1.5$\pm$0.7 & 1.2$\pm$0.7 & -6.8$\pm$1.1 & -4.3$\pm$0.6 \\
Star 3 & 28.8 & 6.3$\pm$0.8 & 5.4$\pm$0.3 & -2.0$\pm$1.0 & 0.8$\pm$0.7 & -7.2$\pm$1.2 & -5.1$\pm$0.9 \\
Star 8 & 14.6 & 6.1$\pm$0.6 & 4.6$\pm$0.3 & -1.8$\pm$0.7 & 0.9$\pm$0.6 & -7.0$\pm$0.9 & -3.8$\pm$0.4 \\
Star 9 & 16.7 & 7.4$\pm$1.8 & 5.5$\pm$0.4 & -2.5$\pm$2.2 & -1.0$\pm$2.5 & -7.2$\pm$2.3 & -4.8$\pm$0.5 \\
\hline
\end{tabular}
\end{table*}

\section{Concluding Remarks}

In the previous sections we characterised the stellar content
associated with the massive protostellar object candidate, IRAS
19343+2026. The stellar content is composed of at least four young
($\sim$ 10$^5$ yr) B type stars and a rich population of low mass
stars at $\sim$ 1- 3 Myr. The FIR peak is modelled as a
7.6 \msun\ massive young star. The cluster is surrounded by a bright
infrared nebula seen from 3--24 $\mu$m, indicative of dust and 
polycyclic aromatic hydrocarbons (PAH)
emission. While the {\em Spitzer} IRAC and MIPS bands display the
infrared nebula that becomes increasingly brighter from 3.6 $\mu$m to
24 $\mu$m, the NIR bands predominantly display the dense stellar
population embedded in this nebula. The following facts can explain
why the cluster population is better detected in deep $K$ band images
than in {\em Spitzer} images. 
The contribution of stellar photospheres rapidly falls off longward of
2 $\mu$m. Further, the background emission due to PAHs is strong in
the bands longward of 4.5 $\mu$m. The above two facts, together with 
the relatively large PSF of the {\em Spitzer} data, 
reduces the sensitivity and contrast of the revealed low mass stars. 
However, the massive stars and the most embedded objects
which appear bright at longer wavelengths are best revealed by the
{\em Spitzer} images. This can explain why deeper NIR $K$-band data 
such as the one presented here can unveil the lower mass population 
better than the GLIMPSE survey.

The fainter and relatively older low mass population revealed by the
NIR images are uniformly arranged over the infrared nebula (see
Fig. 1), coinciding well with the warm dust distribution seen in 24
$\mu$m emission. As shown in Section 3.1, this population is best
described to be 1 - 3 Myr old. In contrast, the four
bright sources are found to be in the age range of 10$^4$ --
10$^5$ yr (see Table 2). This suggests that low mass star formation
occured in the cluster prior to the formation of massive
stars. Although this idea has been argued before, from observations of
other regions \citep[e.g.][]{kum04}, we arrive at the same
conclusion based on a more rigorous analysis using a CM diagram,
$HK$-band spectra, and SED modeling. If all embedded clusters are
born with a universal IMF, then sampling the IMF at any given time,
uniformly over all mass ranges, is naturally expected to result in
fewer massive stars and a greater number of low mass stars. 
Consequently, in the evolution of an embedded cluster (expected to
follow an universal IMF), there exists an early phase when the total
number of massive stars will be zero with a non-zero population
of low mass stars. This simple statistical reasoning suggests that
massive stars should appear after the low mass stars in a young
cluster. Thus, it
appears quite clear that the sequence for star formation is ``low mass
stars first and massive stars next'', at least for clusters forming B
type stars. The situation for clusters forming O type stars requires
further investigation.

\section{Acknowledgments}
We thank the referee for useful comments that improved this
manuscript. We are grateful to Stefan Schmeja for computing the 
nearest-neighbour densities from our photometric data that is 
used in Fig. 2.
This research has received funding from the European
Communities Seventh Framework Programme (FP7/2007-2013) under grant
agreement SF-WF-MSF-230843. The United Kingdom Infrared Telescope is
operated by the Joint Astronomy Centre on behalf of the U.K. Particle
Physics and Astronomy Research Council. This work is based on data 
obtained as part of the UKIRT Infrared Deep Sky Survey.
Kumar is supported by a Ci\^encia 2007 contract, funded by FCT/MCTES (Portugal) and POPH/FSE (EC).
This research made use of data products from the {\em Spitzer} Space
Telescope GLIMPSE legacy survey. These data products are provided by
the services of the Infrared Science Archive operated by the Infrared
Processing and Analysis Center/California Institute of Technology,
funded by the National Aeronautics and Space Administration and the
National Science Foundation. We thank the staff of IAO, Hanle and CREST, 
Hosakote, that made these obervations possible. The facilities at IAO and 
CREST are operated by the Indian Institute of Astrophysics, Bangalore.
We would like to thank A.K. Pandey and M.R. Samal for helpful 
comments and discussion.

\appendix

\section[h]{UFTI $JHK$ photometric data}
Only a few lines of Table A1 are printed here. The complete form of the
table is available online only.
\begin{center}
\begin{table*}
\begin{centering}
\caption{UFTI $JHK$ photometric data of the stars in the region of
IRAS 19343+2026. The complete table is available in electronic form only.}
\label{jhk-catalog}
\begin{tabular}{@{}ccccc}
\hline
RA (2000) & Dec. (2000) & $J$ & $H$ & $K$ \\  
deg & deg & mag & mag & mag \\
\hline
 294.11121 &  20.55862 &  17.58 &  16.83 &  15.11\\
 294.11145 &  20.56095 &  18.24 &  16.05 &  14.88\\
 294.11163 &  20.53695 &  18.10 &  15.80 &  13.94\\
 294.11176 &  20.55444 &  18.14 &  15.73 &  14.39\\
 294.11270 &  20.56126 &  14.75 &  14.36 &  13.83\\
 294.11313 &  20.54862 &  18.49 &  17.86 &  17.57\\
 294.11334 &  20.54917 &  16.44 &  15.40 &  14.75\\
 294.11356 &  20.54623 &  18.13 &  16.56 &  15.53\\
 294.11362 &  20.53854 &  17.95 &  17.26 &  16.64\\
 294.11411 &  20.53953 &  17.85 &  17.20 &  16.81\\
\hline
\end{tabular}
\end{centering}
\end{table*}
\end{center}

\label{lastpage}
\end{document}